\documentclass[journal]{IEEEtran}

\usepackage{listings}
\usepackage{graphicx}
\ifCLASSINFOpdf
\else
\fi

\begin{document}
%
\title{Usage of Permissioned Blockchain Architecture\\ for Big Data in Electronic Medical Records}
%
%
%

\author{Projjal~Gupta,~\IEEEmembership{Member,~IEEE}

Electronics and Communication Engineering

SRM Institute of Science and Technology, Kattankulathur

Email : projjal.gupta@nextech.io

}

\maketitle

\begin{abstract}
With the advent of blockchain technology, multiple research avenues and platforms for dialogue have opened up. However technology transfer to the pubic has not been implemented, such that regular public can access and make use of secure and decentralized software. Most blockchain solutions till date deal with financial applications or monetary transactions, which may not be helpful or be accessible to the general public, especially the lower levels of the financial society. Medi-Chain is a people-first medical blockchain with a usable desktop application and interface which makes use of cutting-edge blockchain technology along with BFT consensus protocols to ensure highly secure and private medical data records. This paper aims to bring about a change in how blockchains-as-a-service is perceived and how adoption of new technology is largely based on usability and ease of adoption.
\end{abstract}


%
\IEEEpeerreviewmaketitle

\section{Introduction}
%
%
%
%
Adoption of any new technology is determined by the usability and efficiency of solving any pre-existing problem, or by replacing any obsolete technological practice, which promote lower performance or lower accountability. Blockchain is one such technological advance which shows a promise of ending the ephemeral issue of privacy and accountability of data. Optimal solutions for privacy and accountability is completely based on trust, which cannot be mitigated by inclusion of a central authority or committee, as it empowers the said central party to own or control the given data, which may lead to misuse or breach of protocol due to polarization.

This paper presents a thorough study of multiple blockchain solutions and centralized system solutions which can be used to implement a logical and working Electronic medical records solutions, while taking an astute approach towards development of said software solution and the ethical and social issues and implications it might pose on public and private sectors of healthcare. Multiple other factors are going to be taken into consideration for the choice of architecture, such as Latency, Scalability, Simplicity, Adoptibility and Data Integrity.

\section{Overview}

Medi-Chain is a blockchain based technology which leverages a Byzantine Fault Tolerant System with state replication over multiple nodes property. This is possible by using BigChainDB as the basis for the technology. Medi-Chain contains of two parts, the blockchain itself and the desktop application built around it to support adoption of the new technology. The Medi-Chain application is built in such a way that it takes public into account and encompasses a multitude of design components to mask the entire backend operation by providing a level of abstraction to the users. To maintain access control, the blockchain contains a separate segment for access control list (acl). The ACL contains information about each and every login details of doctors and administrative data, which controls the login page of the application. The blockchain holds patient details, and provide extended privacy features to all the patients present on the blockchain.

\subsection{Choice of Blockchain Architecture}
Blockchain solutions such as Bitcoin and Ethereum have existed for quite some time, and other solutions such as Hyperledger, Corda R3, BigchainDB, OrbitDB and Quorum are also present on the market. Now, to accomodate data as big as the medical data of multiple patients of a single hospital itself is a huge task. Considering a city or and entire country requires a technolgical solutions engineered to handle such a load. While solutions can't be slow or highly expensive, options like Ethereum and Corda were never apt solutions to begin with. Narrowing the list down, the blockchain requires to have a functional database management solution built in, and data must be sotred in formats which can ease the entire process. 

BigchainDB fits all the criteria above mentioned. Transaction processing speed is a major factor, and BigchainDB delivers on that promise. BigchainDB provides an option to store data in the form of BSON data type, just like in mongo database, and BSON, just like JSON, allows nesting of object to create multiple data entries into a single entity or object. This makes the entire process of read and write sufficiently fast and the data types are quite functional when used in the medical data records area. Salient features of this architecture choice is given below :


\subsubsection{Transaction Processing Speed}
BigchainDB claims to process 320 transactions per second, and can have maximum process speed of 1 Million transactions in 26 Minutes. This performance metric gives it an edge over Quorum and Hyperledger Blockchains.

\subsubsection{Data Structuring}
BigchainDB contains BSON objects, which is like javascript object notation (JSON), but has added support to store timestamps, dates and binary objects and has a faster encoder, which support the transaction speed. Another feature of BSON objects is that while querying operations, irrelevant data is skipped to increase the speed of get operations.

\subsubsection{Fault Tolerance and State Replecation}
BigchainDB is based on Tendermint consensus protocol. This allows the blockchain solution to be fault tolerant upto 33.3\% of all the nodes. This means that if a third of the network is disabled, then the entire network is pulled down such that no state tampering can take place by taking control of the entire network.

\subsubsection{Financial Barrier}
Ethereum blockchain service and Neo Blockchain are expensive to use, in the sense that a transaction fees is deducted for every setter function call on the blockchain. Considering the rising cryptocurrency economy, cost per transactions are at an average ~2-3\$ per new entry. Now in a city setting, it is not a high amount, but in a general country-wise financial status this external cost is not affordable for a country like India. BigchainDB does not require transaction fees just like Hyperledger blockchain service.

\begin{figure}
    \centering
    \includegraphics[width=7cm, height=5cm]{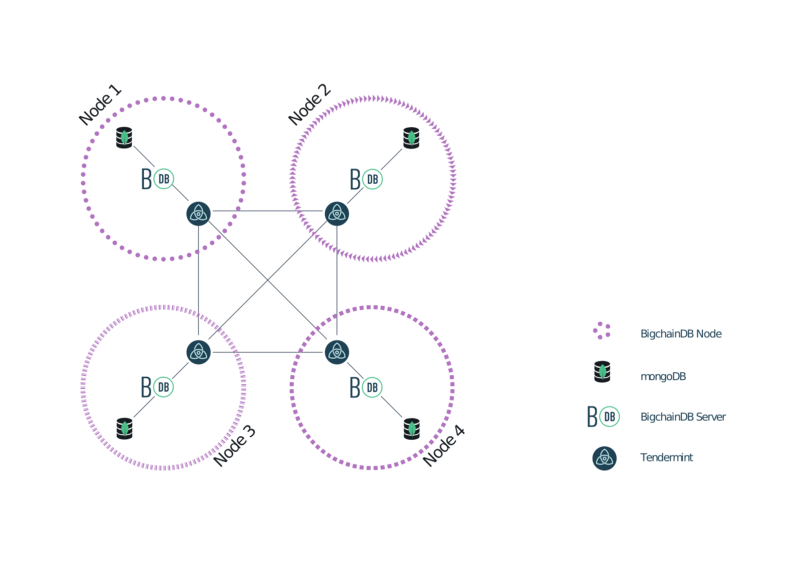}
    \caption{Architecture of a BigchainDB blockchain}
    \label{fig:my_label}
\end{figure}

\subsection{Choice of Front-End Service}
Medichain is aimed towards solving two main problems of blockchain based decentralized applications (Dapps), first being the issue related to the ease of adoption and the second being the problem of finances regarding the cost of each transaction. 

\subsubsection{Ease of Adoption}
Software solutions present today are designed in ways such that the user get the maximum functionality and seamless experience to achieve required tasks. Medi-Chain Desktop application presents an array of options to the user on its dashboard, and presents a vibrant and user friendly user interface to the general public. One does not need to learn about or know the inner workings of the application to use it effectively.

\subsubsection{Application Platform}
Medi-Chain is built on ElectronJS framework, used to build completely seamless native desktop applications. This has an edge over traditional web applications, as a desktop application can provide an interface for a hardware implementation of a kiosk or self-service medical record booth. ElectronJS is highly useful due to its cross-platform development options, and any clinic, hospital and practicing doctor can use it on any computer.

\section{Functions and Structure}

Medi-Chain blockchain stores data in BSON formatted data. However, each data object has a pre-defined structure which needs to be followed so that it can be added to the blockchain. The pre-defined structures are built according to their use cases, and are made sure to not contain any wasteful metadata.

\subsection{Patient Data Structure}
Patient data which is contained in the blockchain is structured using basic information and biodata of the patient, and also contains other details such as mobile number, prescription history, allergies and insurance details as shown below:
\vspace{4mm} 
\begin{lstlisting}
{
    "dbIdentifier" : "",
    "name" : "",
    "gender" : "",
    "age" : "",
    "dob" : "",
    "phone" : "",
    "photo" : "",
    "bloodgroup" : "",
    "superset" : "",
    "docdetails" : {
        "type" : ""
    },
    "allergies" : "",
    "insurance" : ""

}

\end{lstlisting}
\vspace{4mm} 

The data structure is defined above, and is required to fill in all the details on patients. While all the keys in the structure are self-explanatory, "Superset" key is not. It is used to define elevation levels. Elevation levels are given in the figure below. With each added level, more permissions are added to an account so that they can make changes to the blockchain data.

\subsection{Prescription Data Structure}
While doctors prescribe medication to patient, the blockchain checks for 3 things. First is the elevation level of the doctor, secondly, the keys used by the doctor to sign the transaction and third, the patient data to be linked with the prescription.

\vspace{4mm} 
\begin{lstlisting}
{
    "visitId" : "", //Unique Session ID
    "docname" : "",
    "patientnum" : "",
    "problem" : "",
    "prescription" : "",
    "billamt" : ""
}
\end{lstlisting}
\vspace{4mm} 

While the proportion depicts the population of the sector, and each elevation shows heirarchy of the levels

\begin{figure}
    \includegraphics[width=7cm, height=5cm]{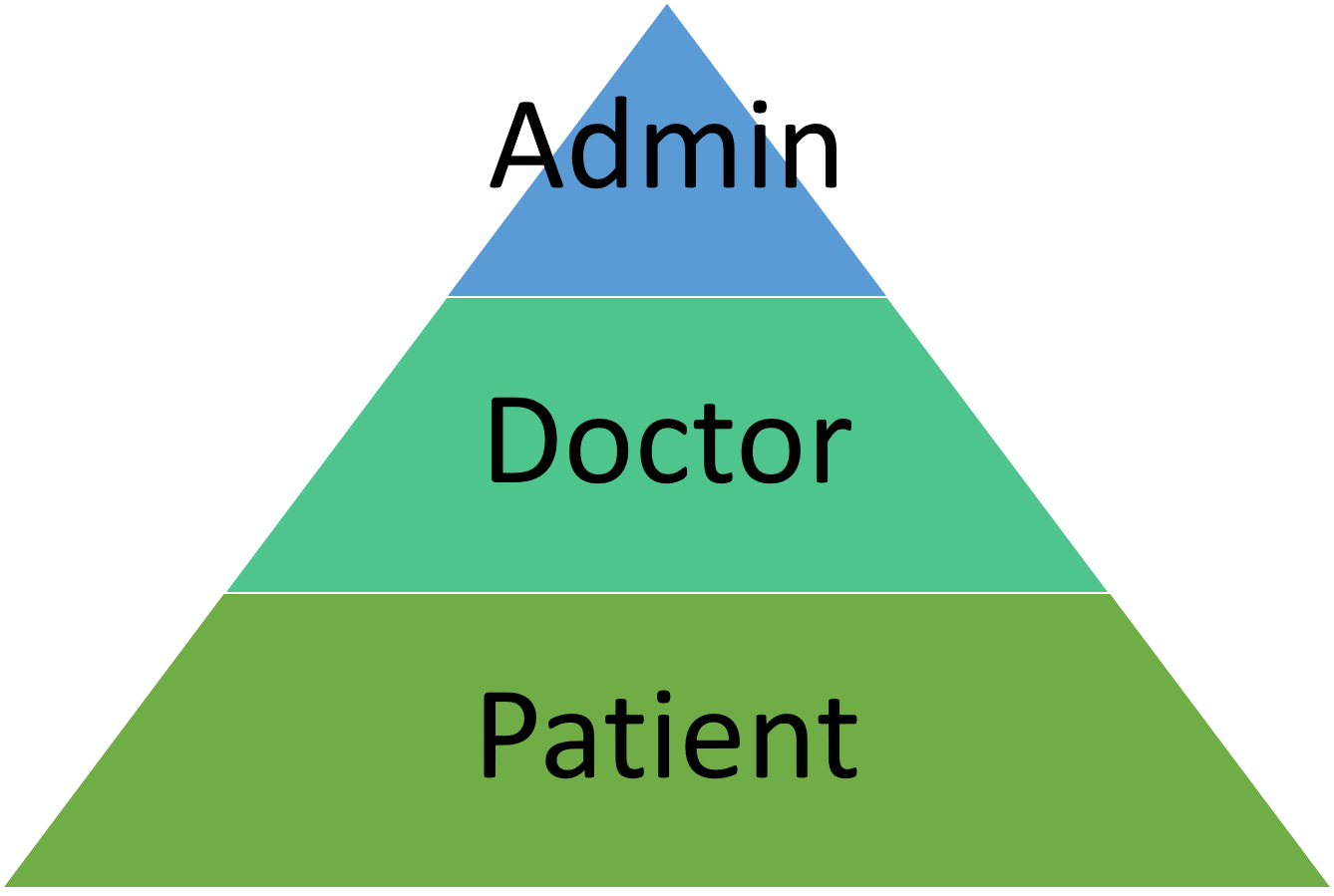}
    \caption{Heirarchial Representation of elevation levels}
\end{figure}

\subsection{Login Data Structure}
The blockchain also saves login data details alongside the patient data. It follows a similar structure to that of the prescription structure.

\vspace{4mm} 
\begin{lstlisting}
{
    "user" : "",
    "pass" : "",
    "mob" : "",
    "superset" : "",
    "key" : ""
}
\end{lstlisting}
\vspace{4mm} 

\subsection{Data Structure Link}
The blockchain contains multiple types of structures and requires to link each and every structure to a single patient. This is possible via linking a single key with each other to form a chain of data. The common key used for chaining of data is the Phone Number of the patient. When a query is made for said mobile number, it can present all the patient data, previous prescriptions and login verification details without any issues.

\vspace{6mm} 
\includegraphics[width=7cm, height=5cm]{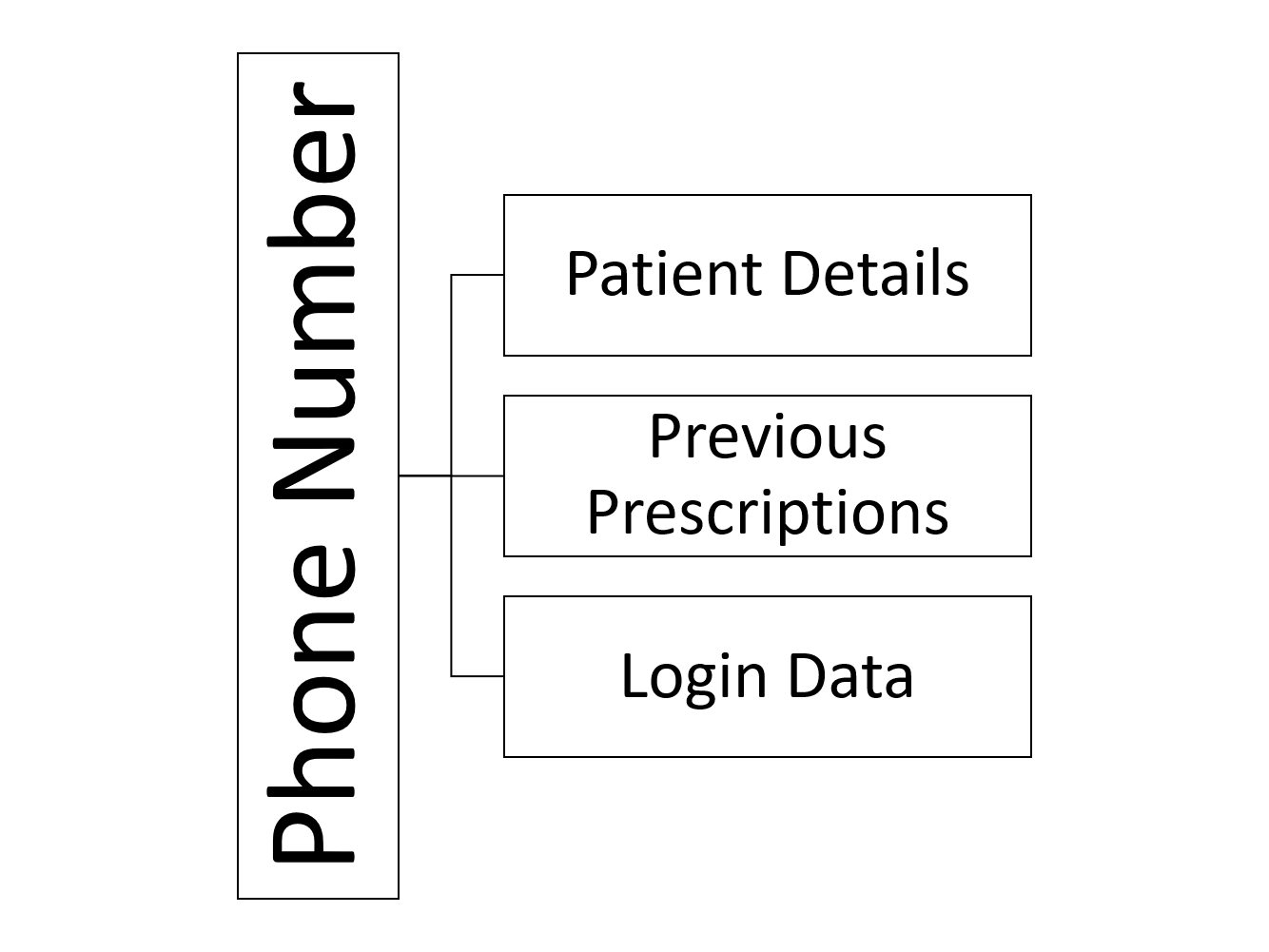}

\section{Additional Features}

Medi-Chain is a blockchain based medical records service, containing millions of records for patients. Big Data of this magnitude can be used to develop a multitude of healthcare frameworks and applications. The medi-chain desktop application encompasses some of the possible use-cases of the big-data of medical records.

\subsection{Open Blood Donation Index}
Donation of blood to help others is noble, and more the number of donors, better are the chances of living for people who desperately require it. Medi-Chain allows doctors to actively search for donors of a certain blood type freely and request for it directly through the blockchain. No personal details are revealed and an automation service present on the blockchain send direct messages to patients for each request.

\subsection{Instant Insurance Claim}
Medi-Chain has open integrations and tie-ups with e-Insurance companies and different companies, such that insurance data can be directly fetched up for each person and billing and claims can be made by the doctor and the patients with one button. However, as insurance claims can be fraudulent, an admin from these companies can review and revoke claims as required.

\subsection{Open Research Data Index}
Research data in the medical fields are scarce and require patients to give consent to fair use. One of the more important issue is privacy of people. Medi-Chain Desktop application can provide a layer of abstraction to hide all personal details such as mobile number, name and any other identifier document. While indexing, one can query for an age-group between some constraints and get a valid number of patient data, without personal details. This enables researchers to work on realtime data stored on the blockchain and get required information directly from the population.

\subsection{IPFS based Prescription Data}
Prescriptions are being hand written for quite a long time. Medi-chain makes sure that conversion to digital data is an option. However to accomodate these old documents, Medichain leverages IPFS or Inter-Planetary File System, which is a open decentralized storage platform. A fork of IPFS with hidden permissions and end-points can be setup to save sensitive information and documents in the form of pdf, png or jpg file types.

\section{Deployment}
Medichain is a public-first medical blockchain service aimed towards making lives better. To host Medi-Chain, one requires to run a minimum of 4 servers to allow the working of the blockchain. However, to maintain such a large database, a suggestion for adoption will be to replace private data banks from hospitals with open Medi-Chain facilities so that data can freely flow amongst all medical practices. 

Private hospitals can easily afford the low-cost blockchain solution. However for government hospitals, a raspberry pi device can be used as a mainframe server. It can host more than 20 screens to actively run on any mobile device per server. On doing a cost breakdown, we fidn that it is highly feasable and can benefit people from all sectors of the society.

\section{Conclusion}
Medichain is a one-of-its-kind medical blockchain solution built to serve and help general public of the country of India. It has multiple feature and can be an anchor point for the start of digital bio-technology revolution. An initiative of such sort and magnitude can only bring blockchain technology closer to the public and enable technology transfer.


%
\appendices

\section*{Acknowledgment}

The author would like to thank  Dr.  C.  Muthamizhchelvan,(Director E\&T) for providing the opportunity,  Dr. B. Neppolian and Dr. S. V.Kasmir Raja, (Dean  Research)  along with their entire team for organizing SRM Research Day. The author gratefully  acknowledges the support of NVIDIA Corporation with their generous donation of the GPUs that were used for this research.

\ifCLASSOPTIONcaptionsoff
  \newpage
\fi


\begin{thebibliography}{1}

\bibitem{IEEEhowto:kopka}
McConaghy, Trent, et al. "BigchainDB: a scalable blockchain database." white paper, BigChainDB (2016).

\bibitem{IEEEhowto:kopka}
Benet, Juan. "Ipfs-content addressed, versioned, p2p file system." arXiv preprint arXiv:1407.3561 (2014).

\bibitem{IEEEhowto:kopka}
Evans, Jae A. "Electronic medical records system." U.S. Patent No. 5,924,074. 13 Jul. 1999.

\bibitem{IEEEhowto:kopka}
Miller, Robert H., and Ida Sim. "Physicians’ use of electronic medical records: barriers and solutions." Health affairs 23.2 (2004): 116-126.

\bibitem{IEEEhowto:kopka}
Azaria, Asaph, et al. "Medrec: Using blockchain for medical data access and permission management." 2016 2nd International Conference on Open and Big Data (OBD). IEEE, 2016.

\bibitem{IEEEhowto:kopka}
Xia, Qi, et al. "BBDS: Blockchain-based data sharing for electronic medical records in cloud environments." Information 8.2 (2017): 44.

\bibitem{IEEEhowto:kopka}
Liu, Paul Tak Shing. "Medical record system using blockchain, big data and tokenization." International conference on information and communications security. Springer, Cham, 2016.

\bibitem{IEEEhowto:kopka}
Kim, Tai-hoon, and Seung-youn Lee. "Security evaluation targets for enhancement of IT systems assurance." International Conference on Computational Science and Its Applications. Springer, Berlin, Heidelberg, 2005.

\bibitem{IEEEhowto:kopka}
Ali, Muhammad Salek, Koustabh Dolui, and Fabio Antonelli. "IoT data privacy via blockchains and IPFS." Proceedings of the Seventh International Conference on the Internet of Things. ACM, 2017.

\bibitem{IEEEhowto:kopka}
Gupta, Projjal, et al. "Smart work-assisting gear." 2018 IEEE 4th World Forum on Internet of Things (WF-IoT). IEEE, 2018.

\bibitem{IEEEhowto:kopka}
Allen, Anita. "Privacy and medicine." (2009).

\bibitem{IEEEhowto:kopka}
Barrows Jr, Randolph C., and Paul D. Clayton. "Privacy, confidentiality, and electronic medical records." Journal of the American medical informatics association 3.2 (1996): 139-148.

\bibitem{IEEEhowto:kopka}
Benaloh, Josh, et al. "Patient controlled encryption: ensuring privacy of electronic medical records." Proceedings of the 2009 ACM workshop on Cloud computing security. ACM, 2009.

\bibitem{IEEEhowto:kopka}
Baumer, David, Julia Brande Earp, and Fay Cobb Payton. "Privacy of medical records: IT implications of HIPAA." ACM SIGCAS Computers and Society 30.4 (2000): 40-47.

\bibitem{IEEEhowto:kopka}
Perera, Gihan, et al. "Views on health information sharing and privacy from primary care practices using electronic medical records." International journal of medical informatics 80.2 (2011): 94-101.

\bibitem{IEEEhowto:kopka}
Damschroder, Laura J., et al. "Patients, privacy and trust: patients’ willingness to allow researchers to access their medical records." Social science \& medicine 64.1 (2007): 223-235.

\bibitem{IEEEhowto:kopka}
Agrawal, Anurag, Jaijit Bhattacharya, Nishant Baranwal, Sushil Bhatla, Salil Dube, Viren Sardana, Devender R. Gaur, Danica Balazova, and Samir K. Brahmachari. "Integrating health care delivery and data collection in rural India using a rapidly deployable eHealth center." PLoS medicine 10, no. 6 (2013): e1001468.

\bibitem{IEEEhowto:kopka}
Scholl, Jeremiah, Shabbir Syed-Abdul, and Luai Awad Ahmed. "A case study of an EMR system at a large hospital in India: challenges and strategies for successful adoption." Journal of biomedical informatics 44.6 (2011): 958-967.

\bibitem{IEEEhowto:kopka}
Lavin, Marianne, and Michael Nathan. "System and method for managing patient medical records." U.S. Patent No. 5,772,585. 30 Jun. 1998.

\bibitem{IEEEhowto:kopka}
Krieger, Nancy. "Overcoming the absence of socioeconomic data in medical records: validation and application of a census-based methodology." American journal of public health 82.5 (1992): 703-710.

\bibitem{IEEEhowto:kopka}
Bates, David W., et al. "A proposal for electronic medical records in US primary care." Journal of the American Medical Informatics Association 10.1 (2003): 1-10.

\bibitem{IEEEhowto:kopka}
Melton III, L. Joseph. "The threat to medical-records research." (1997): 1466-1470.


\end{thebibliography}
\end{document}